\documentclass{ws-ijmpa}
\usepackage{amssymb,amsmath,amsfonts}
\usepackage{graphicx}
\usepackage{graphics}
\usepackage{eepic,epsfig}

\begin{document}

\markboth{A. A. Saharian} {Casimir effect in toroidally
compactified dS spacetime}

%
\catchline{}{}{}{}{} %

\title{CASIMIR EFFECT IN TOROIDALLY COMPACTIFIED DE SITTER SPACETIME}

\author{A. A. Saharian}

\address{Department of Physics, Yerevan State University, 1 Alex Manoogian Street,\\ 0025 Yerevan, Armenia\\
and\\
Departamento de F\'{\i}sica-CCEN, Universidade Federal da
Para\'{\i}ba,\\ 58.059-970, Caixa Postal 5.008, Jo\~{a}o Pessoa,
PB,
Brazil \\
saharian@ictp.it}

\maketitle

\begin{history}
\received{Day Month Year} \revised{Day Month Year}
\end{history}

\begin{abstract}
Vacuum energy density and stresses are investigated  for a scalar
field with general curvature coupling parameter in
$(D+1)$-dimensional de Sitter spacetime with an arbitrary number
of toroidally compactified spatial dimensions. The corresponding
expectation values are presented in the form of the sum of the
vacuum expectation values in uncompactified dS spacetime and the
part induced by the non-trivial topology. In the early stages of
the cosmological evolution the topological parts dominate. In this
limit the behavior of the Casimir densities does not depend on the
curvature coupling parameter and coincides with that for a
conformally coupled massless field. At late stages of the
cosmological expansion the expectation values are dominated by the
part corresponding to uncompactified dS spacetime. The vanishing
of the topological parts is monotonic or oscillatory in dependence
of the mass and the curvature coupling parameter of the field.

\end{abstract}

\ccode{PACS numbers: 04.62.+v, 04.50.-h, 11.10.Kk, 04.20.Gz}

\section{Introduction}

Many of high-energy theories of fundamental physics are formulated
in higher-dimensional spacetimes. In particular, the idea of extra
dimensions has been extensively used in supergravity and
superstring theories. It is commonly assumed that the extra
dimensions are compactified. From an inflationary point of view
universes with compact spatial dimensions, under certain
conditions, should be considered a rule rather than an exception
\cite{Lind04}. The models of a compact universe with non-trivial
topology may play an important role by providing proper initial
conditions for inflation. As it was argued in
Refs.~\refcite{McIn04}, there are many reasons to expect that in
string theory the most natural topology for the universe is that
of a flat compact three-manifold.

In topologically non-trivial spaces the periodicity conditions imposed on
possible field configurations change the spectrum of the vacuum fluctuations
and lead to the Casimir-type contributions to the vacuum expectation values
of physical observables (for the topological Casimir effect and its role in
cosmology see Refs.~\refcite{Most97,Bord01}). In the Kaluza-Klein-type
models the Casimir effect has been used as a stabilization mechanism for
moduli fields which parametrize the size and the shape of the extra
dimensions. The Casimir energy can also serve as a model for dark energy
needed for the explanation of the present accelerated expansion of the
universe (see Refs.~\refcite{Milt03} and references therein). In the present
talk, based on Refs.~\refcite{Saha07,Bell08}, we describe the effects of the
toroidal compactification of spatial dimensions in dS spacetime on the
properties of quantum vacuum for a scalar field with general curvature
coupling parameter. The one-loop quantum effects for a fermionic field on
background of dS spacetime with spatial topology $\mathrm{R}^{p}\times (%
\mathrm{S}^{1})^{q}$ are studied in Refs.~\refcite{Saha08,Beze08}.

The paper is organized as follows. In the next section we describe the
background geometry and present the complete set of eigenfunctions. In
section \ref{sec:vevEMT2} these eigenfunctions are used to evaluate the
mode-sum for the vacuum expectation value of the energy-momentum tensor. The
main results are summarized in section \ref{sec:Conc}.

\section{Background geometry and the eigenfunctions}

\label{sec:Geom}

As a background geometry we consider $(D+1)$-dimensional de Sitter spacetime
($\mathrm{dS}_{D+1}$) generated by a positive cosmological constant $\Lambda
$. In planar (inflationary) coordinates the corresponding line element has
the form
\begin{equation}
ds^{2}=dt^{2}-e^{2t/\alpha }\sum_{l=1}^{D}(dz^{l})^{2},  \label{ds2deSit}
\end{equation}%
with the parameter $\alpha ^{2}=D(D-1)/(2\Lambda )$. We will assume that the
spatial coordinate $z^{l}$, $l=p+1,\ldots ,D$, is compactified to $\mathrm{%
S}^{1}$ of the length $L_{l}$: $0\leqslant z^{l}\leqslant L_{l}$, and for
the other coordinates we have $-\infty <z^{l}<+\infty $, $l=1,\ldots ,p$.
Hence, we consider the spatial topology $\mathrm{R}^{p}\times (\mathrm{S}%
^{1})^{q}$, where $q=D-p$.

This paper is concerned with the scalar vacuum densities induced
by the non-trivial spatial topology. We will consider a free
scalar field with curvature coupling parameter $\xi $. The
corresponding field equation has the form
\begin{equation}
\left( \nabla _{l}\nabla ^{l}+m^{2}+\xi R\right) \varphi =0,  \label{fieldeq}
\end{equation}%
where $R=D(D+1)/\alpha ^{2}$ is the Ricci scalar for $\mathrm{dS}_{D+1}$ and
$\xi $ is the curvature coupling parameter. Let $\mathbf{z}%
_{p}=(z^{1},\ldots ,z^{p})$ and $\mathbf{z}_{q}=(z^{p+1},\ldots ,z^{D})$ be
the position vectors along the uncompactified and compactified dimensions
respectively. We have the following boundary condition along the
compactified dimensions%
\begin{equation}
\varphi (t,\mathbf{z}_{p},\mathbf{z}_{q}+L_{l}\mathbf{e}_{l})=\pm \varphi (t,%
\mathbf{z}_{p},\mathbf{z}_{q}),  \label{periodicBC}
\end{equation}%
where $l=p+1,\ldots ,D$, upper/lower sign corresponds to untwisted/twisted
scalar field, and $\mathbf{e}_{l}$ is the unit vector along the direction of
the coordinate $z^{l}$.

In order to evaluate the vacuum expectation value (VEV) of the the
energy-momentum tensor we will use the direct mode-summation
procedure assuming that the field is prepared in the Bunch-Davies
vacuum state. The corresponding eigenfunctions have the form
\begin{equation}
\varphi _{\sigma }(x)=\left[ \frac{e^{i(\nu -\nu ^{\ast })\pi /2}\eta ^{D}}{%
2^{p+2}\pi ^{p-1}V_{q}\alpha ^{D-1}}\right] ^{1/2}H_{\nu }^{(1)}(k\eta )e^{i%
\mathbf{k}_{p}\cdot \mathbf{z}_{p}+i\mathbf{k}_{q}\cdot \mathbf{z}%
_{q}},\;\eta =\alpha e^{-t/\alpha },  \label{eigfuncD}
\end{equation}%
where $H_{\nu }^{(1)}(x)$ is the Hankel function of the order
\begin{equation}
\nu =\left[ D^{2}/4-D(D+1)\xi -m^{2}\alpha ^{2}\right] ^{1/2},  \label{knD}
\end{equation}%
$V_{q}=L_{p+1}\cdots L_{D}$ is the volume of the compactified subspace, and
\begin{eqnarray}
\mathbf{k}_{p} &=&(k_{1},\ldots ,k_{p}),\;\mathbf{k}_{q}=(k_{p+1},\ldots
,k_{D}),\;k=\sqrt{\mathbf{k}_{p}^{2}+\mathbf{k}_{q}^{2}},\;  \notag \\
\;k_{l} &=&2\pi (n_{l}+g_{l})/L_{l},\;n_{l}=0,\pm 1,\pm 2,\ldots
,\;l=p+1,\ldots ,D.  \label{kD1D2}
\end{eqnarray}%
In (\ref{kD1D2}), $g_{l}=0$ for untwisted scalar and $g_{l}=1/2$ for twisted
scalar field.

\section{Vacuum energy-momentum tensor}

\label{sec:vevEMT2}

In this section we investigate the VEV for the energy-momentum tensor of a
scalar field in $\mathrm{dS}_{D+1}$ with toroidally compactified $q$%
-dimensional subspace. This quantity acts as a source of gravity in the
semiclassical Einstein equations and plays an important role in modelling
self-consistent dynamics involving the gravitational field. Having the
complete set of eigenfunctions we can evaluate the vacuum energy-momentum
tensor by using the mode-sum formula%
\begin{equation}
\langle T_{ik}\rangle _{p,q}=\sum_{\sigma }T_{ik}\{\varphi _{\sigma
}(x),\varphi _{\sigma }^{\ast }(x)\},  \label{EMTsum}
\end{equation}%
where the bilinear form $T_{ik}\{f,g\}$ is determined by the form
of the classical energy-momentum tensor for a scalar field. In the
problem under
consideration the set of quantum numbers $\sigma $ is specified to $(\mathbf{%
k}_{p},\mathbf{n}_{q})$ with $\mathbf{n}_{q}=(n_{p+1},\ldots ,n_{D})$.
Substituting the eigenfunctions (\ref{eigfuncD}) into mode-sum (\ref%
{EMTsum}) and applying to the series over $n_{p+1}$ the Abel-Plana summation
formula (see, for example, Refs.~\refcite{Most97,Saha07Gen}), we find the
following recurrence relation for the VEV of the energy-momentum tensor%
\begin{equation}
\langle T_{i}^{k}\rangle _{p,q}=\langle T_{i}^{k}\rangle _{p+1,q-1}+\Delta
_{p+1}\langle T_{i}^{k}\rangle _{p,q}.  \label{TikDecomp}
\end{equation}%
Here $\langle T_{i}^{k}\rangle _{p+1,q-1}$ is the part corresponding to dS
spacetime with $p+1$ uncompactified and $q-1$ toroidally compactified
dimensions and $\Delta _{p+1}\langle T_{i}^{k}\rangle _{p,q}$ is induced by
the compactness along the $z^{p+1}$ - direction. For the corresponding
energy density one has%
\begin{eqnarray}
\Delta _{p+1}\langle T_{0}^{0}\rangle _{p,q} &=&\frac{2\eta ^{D}L_{p+1}}{%
(2\pi )^{(p+3)/2}V_{q}\alpha ^{D+1}}\sum_{n=1}^{\infty }(\pm 1)^{n}\sum_{%
\mathbf{n}_{q-1}=-\infty }^{+\infty }\int_{0}^{\infty }dx  \notag \\
&&\times \frac{xF_{\nu }^{(0)}(x\eta )}{(nL_{p+1})^{p-1}}f_{(p-1)/2}(nL_{p+1}%
\sqrt{x^{2}+k_{\mathbf{n}_{q-1}}^{2}}),  \label{DelT00}
\end{eqnarray}%
with the notations $\mathbf{n}_{q-1}=(n_{p+2},\ldots ,n_{D})$, $f_{\mu
}(y)=y^{\mu }K_{\mu }(y)$, and%
\begin{eqnarray}
F_{\nu }^{(0)}(y) &=&y^{2}\left[ I_{-\nu }^{\prime }(y)+I_{\nu }^{\prime }(y)%
\right] K_{\nu }^{\prime }(y)+D(1/2-2\xi )y\left[ (I_{-\nu }(y)+I_{\nu
}(y))K_{\nu }(y)\right] ^{\prime }  \notag \\
&&+\left[ I_{-\nu }(y)+I_{\nu }(y)\right] K_{\nu }(y)\left( \nu
^{2}+2m^{2}\alpha ^{2}-y^{2}\right) .  \label{F0}
\end{eqnarray}%
In Eq.~(\ref{DelT00}), the upper/lower sign corresponds to untwisted/twisted
scalar field. The vacuum stresses are presented in the form (no summation
over $i$)%
\begin{eqnarray}
\Delta _{p+1}\langle T_{i}^{i}\rangle _{p,q} &=&A_{p,q}-\frac{4\eta
^{D+2}L_{p+1}}{(2\pi )^{(p+3)/2}V_{q}\alpha ^{D+1}}\sum_{n=1}^{\infty }(\pm
1)^{n}\sum_{\mathbf{n}_{q-1}=-\infty }^{+\infty }\int_{0}^{\infty
}dx\,xK_{\nu }(x\eta )  \notag \\
&&\times \frac{I_{-\nu }(x\eta )+I_{\nu }(x\eta )}{(nL_{p+1})^{p+1}}%
f_{p}^{(i)}(nL_{p+1}\sqrt{x^{2}+k_{\mathbf{n}_{q-1}}^{2}}),  \label{DelTii}
\end{eqnarray}%
where we have introduced the notations%
\begin{eqnarray}
f_{p}^{(i)}(y) &=&f_{(p+1)/2}(y),\;i=1,\ldots ,p,  \notag \\
f_{p}^{(p+1)}(y) &=&-y^{2}f_{(p-1)/2}(y)-pf_{(p+1)/2}(y),  \label{fp+1} \\
f_{p}^{(i)}(y) &=&(nL_{p+1}k_{i})^{2}f_{(p-1)/2}(y),\;i=p+2,\ldots ,D.
\notag
\end{eqnarray}%
The first term on the right of Eq.~(\ref{DelTii}) is given by
\begin{eqnarray}
A_{p,q} &=&\frac{2\eta ^{D}L_{p+1}}{(2\pi )^{(p+3)/2}V_{q}\alpha ^{D+1}}%
\sum_{n=1}^{\infty }(\pm 1)^{n}\sum_{\mathbf{n}_{q-1}=-\infty }^{+\infty
}\int_{0}^{\infty }dx\,\frac{xF_{\nu }(x\eta )}{(nL_{p+1})^{p-1}}  \notag \\
&&\times f_{(p-1)/2}(nL_{p+1}\sqrt{x^{2}+k_{\mathbf{n}_{q-1}}^{2}}),
\label{A}
\end{eqnarray}%
with the notation%
\begin{eqnarray}
F_{\nu }(y) &=&\left[ 2(D+1)\xi -D/2\right] y(\left[ I_{-\nu }(y)+I_{\nu }(y)%
\right] K_{\nu }(y))^{\prime }+\left( 4\xi -1\right) y^{2}K_{\nu }^{\prime
}(y)  \notag \\
&&\times \left[ I_{-\nu }^{\prime }(y)+I_{\nu }^{\prime }(y)\right] +\left[
I_{-\nu }(y)+I_{\nu }(y)\right] K_{\nu }(y)\left[ \left( 4\xi -1\right)
\left( y^{2}+\nu ^{2}\right) \right] .  \label{Fy}
\end{eqnarray}%
As it is seen from the obtained formulae, the topological parts in
the VEVs are time-dependent and, hence, the local dS symmetry is
broken by them. By taking into account the relation between the
conformal and synchronous time coordinates, we see that the VEV of
the energy-momentum tensor is a function of the combinations
$L_{l}/\eta =L_{l}e^{t/\alpha }/\alpha $, which is the comoving
length of the compactified dimension measured in units of the dS
curvature scale.

The recurring application of formula (\ref{TikDecomp}) allows us to write
the VEV in the form%
\begin{equation}
\langle T_{i}^{k}\rangle _{p,q}=\langle T_{i}^{k}\rangle _{\mathrm{dS}%
}+\langle T_{i}^{k}\rangle _{\mathrm{c}},\;\langle T_{i}^{k}\rangle _{%
\mathrm{c}}=\sum_{l=1}^{q}\Delta _{D-l+1}\langle T_{i}^{k}\rangle _{D-l,l},
\label{TikComp}
\end{equation}%
where the part corresponding to uncompactified dS spacetime, $\langle
T_{i}^{k}\rangle _{\mathrm{dS}}$, is explicitly decomposed. The part $%
\langle T_{i}^{k}\rangle _{\mathrm{c}}$ is induced by the comactness of the $%
q$-dimensional subspace. This part is finite and the
renormalization is needed for the uncompactified dS part only. We
could expect this result, since the local geometry is not changed
by the toroidal compactification.

For a conformally coupled massless scalar field $\nu =1/2$ and, by using the
formulae for $I_{\pm 1/2}(x)$ and $K_{1/2}(x)$, after the integration over $x
$ from formulae (\ref{DelT00}), (\ref{DelTii}) we find (no summation over $i$%
)%
\begin{equation}
\Delta _{p+1}\langle T_{i}^{i}\rangle _{p,q}=-\frac{2(\eta /\alpha )^{D+1}}{%
(2\pi )^{p/2+1}V_{q-1}}\sum_{n=1}^{\infty }(\pm 1)^{n}\sum_{\mathbf{n}%
_{q-1}=-\infty }^{+\infty }\frac{g_{p}^{(i)}(nL_{p+1}k_{\mathbf{n}_{q-1}})}{%
(nL_{p+1})^{p+2}},  \label{DelTConf}
\end{equation}%
with the notations%
\begin{eqnarray}
g_{p}^{(0)}(y) &=&g_{p}^{(i)}(y)=f_{p/2+1}(y),\;i=1,\ldots ,p,  \notag \\
g_{p}^{(p+1)}(y) &=&-(p+1)f_{p/2+1}(y)-y^{2}f_{p/2}(y),  \label{gi} \\
g_{p}^{(i)}(y) &=&(nL_{p+1}k_{i})^{2}f_{p/2}(y),\;i=p+2,\ldots ,D.  \notag
\end{eqnarray}%
In this case the topological part in the VEV of the energy-momentum tensor
is traceless and the trace anomaly is contained in the uncompactified dS
part only. Formula (\ref{DelTConf}) could be obtained from the corresponding
result in $(D+1)$-dimensional Minkowski spacetime with spatial topology $%
\mathrm{R}^{p}\times (\mathrm{S}^{1})^{q}$, taking into account that two
problems are conformally related: $\Delta _{p+1}\langle T_{i}^{k}\rangle
_{p,q}=\Delta _{p+1}\langle T_{i}^{k}\rangle _{p,q}^{\mathrm{(M)}%
}/a^{D+1}(\eta )$, where $a(\eta )=\alpha /\eta $ is the scale factor. This
relation is valid for any conformally flat bulk. The similar formula takes
place for the total topological part $\langle T_{i}^{k}\rangle _{\mathrm{c}}$%
. Note that, in this case the expressions for $\Delta _{p+1}\langle
T_{i}^{k}\rangle _{p,q}$ are obtained from the formulae for $\Delta
_{p+1}\langle T_{i}^{k}\rangle _{p,q}^{\mathrm{(M)}}$ replacing the lengths $%
L_{l}$ of the compactified dimensions by the comoving lengths $\alpha
L_{l}/\eta $, $l=p,\ldots ,D$.

Now we turn to the investigation of the topological part in the
VEV of the energy-momentum tensor in the asymptotic regions of the
ratio $L_{p+1}/\eta $. For small values of this ratio,
$L_{p+1}/\eta \ll 1$, to the leading order $\Delta _{p+1}\langle
T_{i}^{k}\rangle _{p,q}$ coincides with the corresponding result
for a conformally coupled massless field, given by
(\ref{DelTConf}).
For fixed value of the ratio $L_{p+1}/\alpha $, this limit corresponds to $%
t\rightarrow -\infty $ and the topological part $\langle T_{i}^{k}\rangle _{%
\mathrm{c}}$ behaves like $\exp [-(D+1)t/\alpha ]$. By taking into
account that the part $\langle T_{i}^{k}\rangle _{\mathrm{dS}}$ is
time independent, from here we conclude that in the early stages
of the cosmological expansion the topological part dominates in
the VEV\ of the energy-momentum tensor. In particular, in this
limit the total energy density is negative.

For small values of the ratio $\eta /L_{p+1}$, we introduce a new
integration variable $y=L_{p+1}x$ and expand the integrand by
using the formulae for the modified Bessel functions for small
arguments. For real
values of the parameter $\nu $ we find%
\begin{eqnarray}
\Delta _{p+1}\langle T_{0}^{0}\rangle _{p,q} &\approx &\frac{2^{\nu }D\left[
D/2-\nu +2\xi \left( 2\nu -D-1\right) \right] }{(2\pi
)^{(p+3)/2}L_{p+1}^{-q}V_{q}\alpha ^{D+1}}\Gamma (\nu )\left( \frac{\eta }{%
L_{p+1}}\right) ^{D-2\nu }  \notag \\
&&\times \sum_{n=1}^{\infty }(\pm 1)^{n}\sum_{\mathbf{n}_{q-1}=-\infty
}^{+\infty }\frac{f_{(p+1)/2-\nu }(nL_{p+1}k_{\mathbf{n}_{q-1}})}{%
n^{(p+1)/2-\nu }}.  \label{T00smallEta}
\end{eqnarray}%
In particular, this quantity is positive for a minimally coupled
scalar field and for a conformally coupled massive scalar field.
For a conformally coupled massless scalar the coefficient in
(\ref{T00smallEta})
vanishes. For the vacuum stresses the second term on the right of formula (%
\ref{DelTii}) is suppressed with respect to the first term given by (\ref{A}%
) by the factor $(\eta /L_{p+1})^{2}$ for $i=1,\ldots ,p+1$, and by the
factor $(\eta k_{i})^{2}$ for $i=p+2,\ldots ,D$. As a result, to the leading
order we have the relation (no summation over $i$)
\begin{equation}
\Delta _{p+1}\langle T_{i}^{i}\rangle _{p,q}\approx (2\nu /D)\Delta
_{p+1}\langle T_{0}^{0}\rangle _{p,q},\;\eta /L_{p+1}\ll 1,
\label{TiismallEta}
\end{equation}%
between the energy density and stresses, $i=1,\ldots ,D$. The
coefficient in this relation does not depend on $p$ and, hence, it
takes place for the total topological part of the VEV as well.
Hence, in the limit under consideration the topological parts in
the vacuum stresses are isotropic. Note that this limit
corresponds to late times in terms of synchronous time coordinate
$t$, $(\alpha /L_{p+1})e^{-t/\alpha }\ll 1$, and the topological
part in the VEV is suppressed by the factor $\exp [-(D-2\nu
)t/\alpha ]$. For a conformally coupled massless scalar field the
coefficient of the leading term vanishes and the topological parts
are suppressed by the factor $\exp [-(D+1)t/\alpha ]$. As the
uncompactified dS part is constant, it dominates the topological
part at the late stages of the cosmological evolution.

For small values of the ratio $\eta /L_{p+1}$ and for purely imaginary $\nu $%
, the energy density behaves like%
\begin{equation}
\Delta _{p+1}\langle T_{0}^{0}\rangle _{p,q}\approx \frac{4De^{-Dt/\alpha }B%
}{(2\pi )^{(p+3)/2}\alpha L_{p+1}^{p}V_{q}}\sin [2|\nu |t/\alpha +2|\nu |\ln
(L_{p+1}/\alpha )+\phi _{0}],  \label{T00ImEta}
\end{equation}%
where the parameters $B$ and $\phi _{0}$ are defined by the relation
\begin{eqnarray}
Be^{i\phi _{0}} &=&2^{i|\nu |}[|\nu |(1/2-2\xi )+i(D/4-(D+1)\xi )]\Gamma
(i|\nu |)  \notag \\
&&\times \sum_{n=1}^{\infty }(\pm 1)^{n}\sum_{\mathbf{n}_{q-1}=-\infty
}^{+\infty }n^{2i|\nu |-p-1}f_{(p+1)/2-i|\nu |}(nL_{p+1}k_{\mathbf{n}%
_{q-1}}).  \label{Bphi0}
\end{eqnarray}%
In the same limit, the main contribution into the vacuum stresses comes from
the term $A_{p,q}$ in (\ref{A}) and one has (no summation over $i$)%
\begin{equation}
\Delta _{p+1}\langle T_{i}^{i}\rangle _{p,q}\approx \frac{8|\nu
|e^{-Dt/\alpha }B}{(2\pi )^{(p+3)/2}\alpha L_{p+1}^{p+1}V_{q-1}}\cos [2|\nu
|t/\alpha +2|\nu |\ln (L_{p+1}/\alpha )+\phi _{0}].  \label{TiiImEta}
\end{equation}%
Hence, in the case under consideration at late stages of the cosmological
evolution the topological part is suppressed by the factor $\exp (-Dt/\alpha
)$ and the damping of the corresponding VEV has an oscillatory nature.

\begin{figure}[pb]
\centerline{\psfig{file=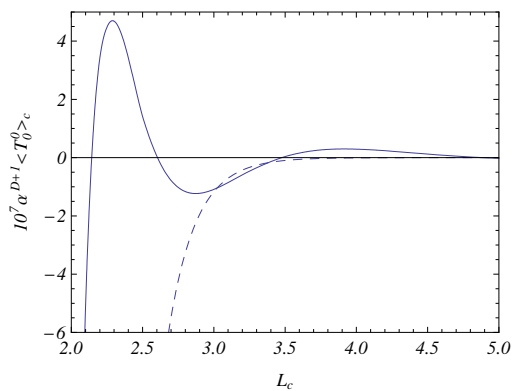,width=6.7cm}} \vspace*{8pt}
\caption{Topological part of the vacuum energy density, $\alpha
^{D+1}\langle T_{0}^{0}\rangle _{\mathrm{c}}$, in
$\mathrm{dS}_{5}$ with topology $\mathrm{R}^{3}\times
\mathrm{S}^{1}$ as a function of the comoving length of the
compactified dimension in units of $\alpha $, $L_{c}=L/\eta $, for
the value of the parameter $\alpha m=4$. The dashed curve presents
the corresponding quantity in Minkowski spacetime with topology $\mathrm{R}%
^{3}\times \mathrm{S}^{1}$ as a function of the length of the
compactified dimension in the same units: $L_{c}=L/\alpha $.
\label{fig1}}
\end{figure}

In the special case of topology $\mathrm{R}^{D-1}\times \mathrm{S}^{1}$ with
the length of the compactified dimension $L_{p+1}=L$, for the topological
part in the energy density we have%
\begin{equation}
\langle T_{0}^{0}\rangle _{\mathrm{c}}=\frac{2(\eta /L)^{D-2}}{(2\pi
)^{D/2+1}\alpha ^{D+1}}\sum_{n=1}^{\infty }\frac{(\pm 1)^{n}}{n^{D-2}}%
\int_{0}^{\infty }dx\,xF_{\nu }^{(0)}(x)f_{D/2-1}(nxL/\eta ).
\label{T00SpS1}
\end{equation}%
We recall that the quantity $L/\eta $ is the comoving length of
the compactified dimension measured in units of the dS curvature
scale $\alpha $. Note that the corresponding quantity in the
Minkowski spacetime with topology $\mathrm{R}^{D-1}\times
\mathrm{S}^{1}$ has the form
\begin{equation}
\langle T_{0}^{0}\rangle _{\mathrm{c}}^{\mathrm{(M)}}=-\frac{2}{(2\pi
)^{(D+1)/2}L^{D+1}}\sum_{n=1}^{\infty }\frac{(\pm 1)^{n}}{n^{D+1}}%
f_{(D+1)/2}(nLm),  \label{T00Mink}
\end{equation}%
and is always positive for an untwisted scalar field. In order to
illustrate the oscillatory behavior, in Fig.~\ref{fig1} by the
full curve we have plotted the topological part in the VEV of the
energy density for an untwisted scalar field in $\mathrm{dS}_{5}$
with topology $\mathrm{R}^{3}\times \mathrm{S}^{1}$ as a function
of the comoving length of the compactified dimension in units of
$\alpha $: $L_{c}=L/\eta $ for the value of the parameter $\alpha
m=4$. This topology corresponds to the original Kaluza-Klein
model. The dashed curve presents
the corresponding quantity in Minkowski spacetime with topology $\mathrm{R}%
^{3}\times \mathrm{S}^{1}$ (formula (\ref{T00Mink}) with $D=4$) as
a function of the length of the compactified dimension in the same
units: $L_{c}=L/\alpha $.

\section{Conclusion}

\label{sec:Conc}

Motivated by the fact that dS spacetime naturally arises in a
number of contexts, in the present paper we consider the Casimir
densities for a massive scalar field in $(D+1)$-dimensional dS
spacetime having the spatial topology $\mathrm{R}^{p}\times
(\mathrm{S}^{1})^{q}$. Both cases of the periodicity and
antiperiodicity conditions along the compactified dimensions are
discussed. We have derived a recurrence relation which presents
the
vacuum energy-momentum tensor for the $\mathrm{dS}_{D+1}$ with topology $%
\mathrm{R}^{p}\times (\mathrm{S}^{1})^{q}$ in the form of the sum of the
energy-momentum tensor for the topology $\mathrm{R}^{p+1}\times (\mathrm{S}%
^{1})^{q-1}$ and the additional part induced by the compactness of the $%
(p+1) $th spatial dimension. The repeated application of the recurrence
formula allows us to present the expectation value of the energy-momentum
tensor as the sum of the uncompactified dS and topological parts. Since the
toroidal compactification does not change the local geometry, in this way
the renormalization of the energy-momentum tensor is reduced to that for
uncompactifeid $\mathrm{dS}_{D+1}$.

At early stages of the cosmological expansion, corresponding to
$t\rightarrow -\infty $, the vacuum energy-momentum tensor
coincides with the corresponding quantity for a conformally
coupled massless field and the topological part behaves like
$e^{-(D+1)t/\alpha }$. In this limit the topological part
dominates in the VEV. At late stages of the cosmological
expansion, $t\rightarrow +\infty $, the behavior of the
topological part depends on the value of $\nu $. For real values
of this parameter the leading term in the corresponding asymptotic
expansion is given by formula (\ref{T00smallEta}) and the vacuum
stresses are isotropic. In this limit the topological part is
suppressed by the factor $e^{-(D-2\nu )t/\alpha }$. In the same
limit and for pure imaginary values of the parameter $\nu $ the
asymptotic behavior of the topological part in the VEV of the
energy-momentum tensor is described by formulae (\ref{T00ImEta}),
(\ref{TiiImEta}) and the topological terms oscillate with the
amplitude going to the zero as $e^{-Dt/\alpha }$ for $t\rightarrow
+\infty $.

\section*{Acknowledgments}

The author acknowledges the organizers of the 7th Alexander
Friedmann International Seminar on Gravitation and Cosmology for a
support. The work has been supported by the Armenian Ministry of
Education and Science Grant No. 119 and by Conselho Nacional de
Desenvolvimento Cient\'{\i}fico e Tecnol\'{o}gico (CNPq).

\end{document}